\documentclass[journal]{IEEEtran}

\pdfoutput=1

\usepackage{amsmath,bm,amssymb,amsfonts,amsthm}
\usepackage{algorithm,algorithmic}
\usepackage{graphicx}
\usepackage{xcolor}
\usepackage{relsize}
\usepackage{textcomp}
\usepackage{listings}

\usepackage{changepage}
\usepackage{hhline}
\usepackage{multirow}
\usepackage{nomencl}
\usepackage{mathtools}
\usepackage{commath}
\usepackage{booktabs}
\usepackage{subfiles}
\usepackage{tabularx}
\usepackage{lscape}
\usepackage{rotating}

\usepackage[normalem]{ulem}
\usepackage[utf8]{inputenc}
\usepackage[hyphens]{url}
\usepackage{lineno}
    \modulolinenumbers[5]
\usepackage[caption=false]{subfig}
\usepackage{enumitem}
    \setlist{nolistsep}
\usepackage{gensymb}
\usepackage[hidelinks]{hyperref}

\usepackage{scalerel}
\usepackage{tikz}
\usetikzlibrary{svg.path}
\definecolor{orcidlogocol}{HTML}{A6CE39}
\tikzset{
  orcidlogo/.pic={
    \fill[orcidlogocol] svg{M256,128c0,70.7-57.3,128-128,128C57.3,256,0,198.7,0,128C0,57.3,57.3,0,128,0C198.7,0,256,57.3,256,128z};
    \fill[white] svg{M86.3,186.2H70.9V79.1h15.4v48.4V186.2z}
                 svg{M108.9,79.1h41.6c39.6,0,57,28.3,57,53.6c0,27.5-21.5,53.6-56.8,53.6h-41.8V79.1z M124.3,172.4h24.5c34.9,0,42.9-26.5,42.9-39.7c0-21.5-13.7-39.7-43.7-39.7h-23.7V172.4z}
                 svg{M88.7,56.8c0,5.5-4.5,10.1-10.1,10.1c-5.6,0-10.1-4.6-10.1-10.1c0-5.6,4.5-10.1,10.1-10.1C84.2,46.7,88.7,51.3,88.7,56.8z};
  }
}
\newcommand\orcidicon[1]{\href{https://orcid.org/#1}{\mbox{\scalerel*{
\begin{tikzpicture}[yscale=-1,transform shape]
\pic{orcidlogo};
\end{tikzpicture}
}{|}}}}

\usepackage[noadjust]{cite}

\begin{document}

\title{\huge The Risk of Hidden Failures to the United States \\ Electrical Grid and Potential for Mitigation}

\author{
    Arthur~K.~Barnes $^{1}$\orcidicon{0000-0001-9718-3197},
    Adam~Mate $^{1}$\orcidicon{0000-0002-5628-6509},
    and Jose~E.~Tabarez $^{1}$\orcidicon{0000-0003-4800-6340}

\thanks{Manuscript submitted: Jul.~15,~2021.
Current version: Oct.~5,~2021.
}

\thanks{$^{1}$ The authors are with the Advanced Network Science Initiative at Los Alamos National Laboratory, Los Alamos, NM 87545 USA. Email:\{abarnes, amate, jtabarez\}@lanl.gov.}

\thanks{LA-UR-21-26802. Approved for public release; distribution is unlimited.}

}

\markboth{IEEE/PES 53rd North American Power Symposium, November~2021}{}

\maketitle


\begin{abstract}
Hidden failures present a noticeable impact to the reliability of the United States electrical grid. These hazards are responsible for protective device misoperations and can cause  multiple-element contingencies across nearby components, greatly increasing the likelihood of cascading events.
This paper provides an in-depth overview of the causes and risks of hidden failures and discusses methods for identifying critical locations where hidden failures could pose a risk of cascading failure, with the ultimate goal being to identify efficient mitigation methods that can prevent their occurrence in protective relays.
\end{abstract}


\section{Introduction} \label{sec:intro}
\indent

Research has shown that despite the recent slowdown in electrical load growth in the North American energy system, the occurrence of large-scale blackouts has not been reduced \cite{hines_trends_2008}.
One contributor to blackouts is the presence of hidden failures in the electrical grid that result in a single fault turning into a multi-element contingency. As described by Elizondo~et~al.~\cite{elizondo_hidden_2001}, a hidden failure is defined as a relay that is misconfigured or faulty such that it will cause the inappropriate removal of system assets during an event (e.g., an out-of-zone fault).
It can be stated that hidden failures in protective relaying present a risk of cascading failures to the entire grid.

Hidden failures are considered to occur when protection activates on account of a misconfiguration or internal failure, resulting in an inappropriate removal of system assets during an event (e.g., an out-of-zone fault) ~\cite{elizondo_hidden_2001, horowitz_boosting_2003}. For the most part, hidden failures are synonymous with misoperations and are a consequence of biasing a protection system towards dependability (i.e., always clearing a fault on the protected element) versus security (i.e., never misoperating when a fault has not occurred on the protected element) \cite{blackburn_protective_2007}.
However, the failure of local protection to operate for an in-zone fault, triggering remote backup protection to remove additional system assets, can also be considered a hidden failure \cite{qiu_risk_2003}. 
Of particular concern are hidden failures that cause the loss of a set of critical assets, resulting in a severe loss of load or a cascading failure event, which in turn could lead to prolonged outages and expensive damage.

While the main cause of power outages remains natural disasters (e.g., extreme weather, forest fires, earthquakes) \cite{hines_trends_2008}, it has been demonstrated that hidden failures present a noticeable impact to the reliability of the United States electrical grid.
A number of blackouts are attributed to hidden failures, including the 1977 New York blackout and the 1995 southern Idaho event (in which a line incorrectly tripped on account of a fault on a parallel line, resulting in a subsequent overload trip of a third parallel line, requiring the shed of several GW loads on account of an underfrequency event) \cite{phadke_expose_1996}.
It is important to mention that hidden failures (including those that cause cascading failures) are responsible for an estimated 10\% rate of misoperations across protective device operations \cite{meliopoulos_dynamic_2017} and 70\% of $n-2$ contingencies are caused by relay misoperations \cite{gao_case_2013}. Additionally, while the electrical grid was designed to be robust to $n-1$ contingencies, hidden failures can cause multiple-element contingencies for which the risk of load loss or cascading events is much higher \cite{qiu_risk_2003}.

The contributions of this paper include the following: Section~\ref{sec:causes} presents a taxonomy of the causes of hidden failures, Section~\ref{sec:identifying} describes methods for identifying critical locations where hidden failures could pose a risk of cascading failure, and Section~\ref{sec:mitigation} describes methods for mitigation of hidden failures.
Although methods in the literature exist for mitigating the impacts of cascading failures -- e.g., remedial action schemes \cite{vaiman_mitigation_2013} or the ``Reciprocal Altruism'' method \cite{hines_cascading_2009} -- this paper specifically focuses on methods to reduce the likelihood of protective misoperations caused by hidden failures.

\section{Causes of Hidden Failures} \label{sec:causes}
\indent

There are several main causes of hidden failures: relay hardware failures, bad electrical connections, incorrect settings, inappropriate design or application of protective schemes, and operator errors \cite{qiu_risk_2003, gao_case_2013, albinali_2016, tamronglak_analysis_1994}.
Defects in protection systems (e.g., current transformers and voltage transformers, wiring, protective relays, communications systems, circuit breakers) can cause a misoperation during external faults, under-voltages, overloads, switching events, transformer energization, or reverse power flows \cite{elizondo_hidden_2001, phadke_expose_1996, fan_dynamic_2015}.

According to Tamronglak~\cite{tamronglak_analysis_1994}, relays that can experience hidden failures include: directional comparison blocking and unblocking relays, permissive overreaching and under-reaching transfer trip relays, Zone~2 and Zone~3 distance relays, directional overcurrent ground relays, and differential relays.
It is notable that many of these relays include pilot protection, a complex protection with an increased number of potential failure modes \cite{barnes21-admitrelay, elizondo_methodology_2003}.

The operation of some common protection schemes and causes of hidden failure for these schemes are discussed below.
\newpage

\vspace{0.1in}
\noindent \textit{Directional Comparison Blocking (DCB) Relays}
\vspace{0.1in}
\indent

In a DCB relaying scheme, the line is protected by two directional distance relays on either end for line-line faults and directional overcurrent relays for line-ground faults \cite{tamronglak_analysis_1994, elizondo_methodology_2003, blackburn_protective_2007}. There is a bidirectional channel along the line by which each relay can send a blocking signal preventing the other from operating.
An important feature of this scheme is that the channel is only transmitting the blocking signal if a relay detects an external fault. It is therefore biased towards dependability in that the breakers on either end of the line can operate if the communications channel is inoperable.

\begin{figure}[!htbp]
\centering
\includegraphics[scale=0.5]{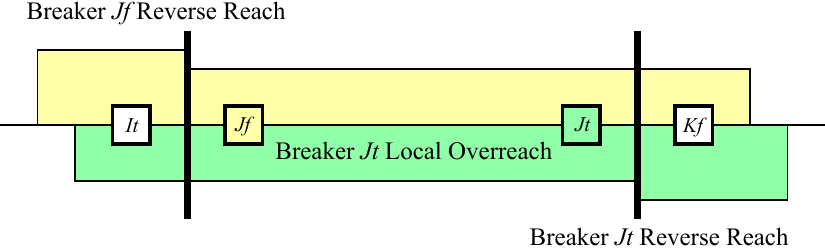}
\caption{Oneline diagram for a DCB relaying scheme.}
\label{fig:dcb-oneline}
\end{figure}

If a relay's directional element detects that the fault is external, it will send a blocking signal to the relay on the other end of the line to prevent it from operating. Should the blocking signal fail to be received, the line will be tripped, likely causing an $n-2$ contingency.
This can occur when: either or both reverse-reach fault detectors fail to detect a fault; or the communications link has failed.

\vspace{0.1in}
\noindent \textit{Directional Comparison Unblocking (DCUB) Relays}
\vspace{0.1in}
\indent

\begin{figure}[!htbp]
\centering
\includegraphics[scale=0.5]{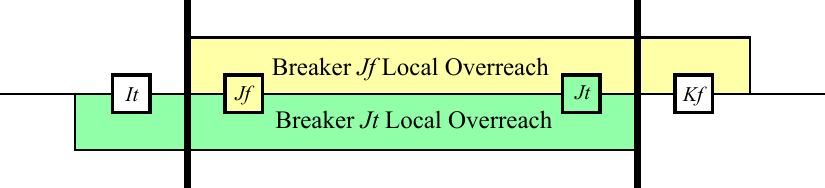}
\caption{Oneline diagram for a DCUB and a POTT relaying scheme.}
\label{fig:dcub-oneline}
\end{figure}

\begin{figure}[!htbp]
\centering
\includegraphics[scale=0.225]{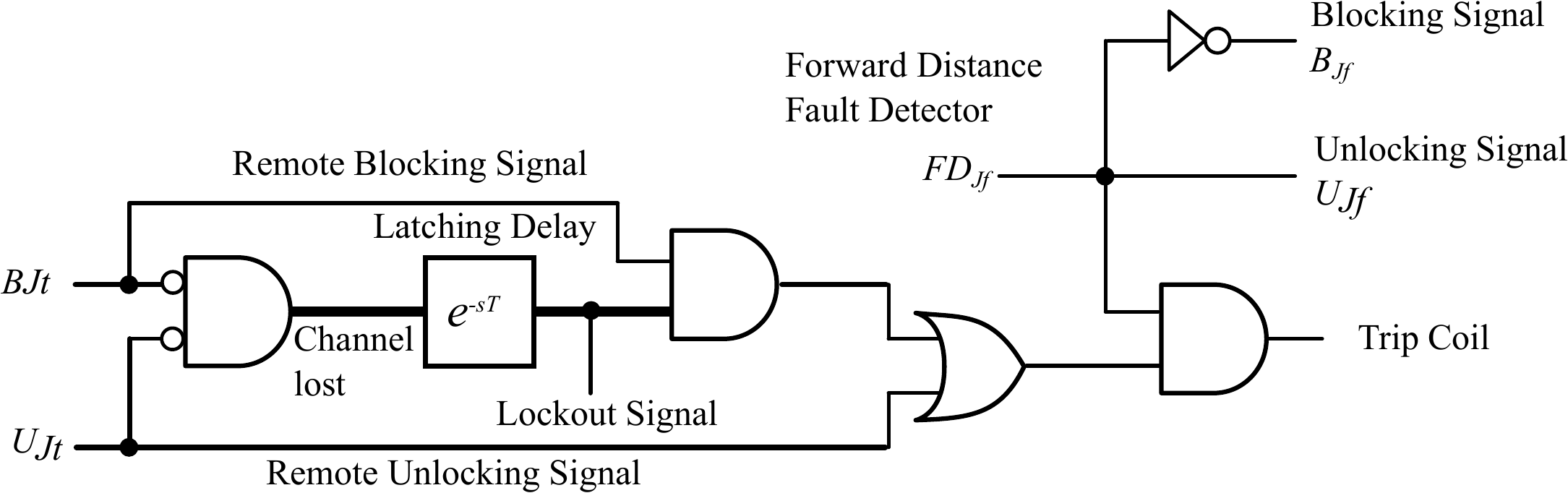}
\caption{Schematic for one relay in a DCUB scheme.}
\label{fig:dcub-logic}
\end{figure}

In a DCUB relaying scheme, the line is again protected by two directional distance relays on either end, as illustrated in Fig.~\ref{fig:dcub-oneline}. Two separate unidirectional communications channels are used to send a continuous blocking signal, typically a tone of a particular frequency \cite{tamronglak_analysis_1994, elizondo_methodology_2003, blackburn_protective_2007}.
If the directional unit detects that a fault is within the protected zone of the relay, it will change the tone to the unblocking frequency. The logic is illustrated in Fig.~\ref{fig:dcub-logic}.

An important benefit over the directional blocking scheme is that it will not experience a hidden failure in the communications channel as it is continuously operating and such a failure will be immediately detected without the presence of a fault.
However, this scheme can experience hidden failures when: either \emph{but not both} of the forward or reverse directional fault detectors continuously report a fault; or either or both receivers continuously report an unblocking signal.

\vspace{0.1in}
\noindent \textit{Permissive Underreaching Transfer Trip (PUTT) Relays}
\vspace{0.1in}
\indent

\begin{figure}[!htbp]
\centering
\includegraphics[scale=0.6]{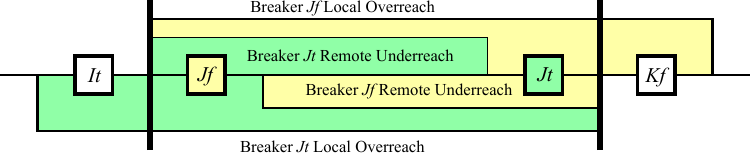}
\caption{Oneline diagram for a PUTT relaying scheme.}
\label{fig:putt-oneline}
\end{figure}

In a PUTT relaying scheme, the protective relays on either end of the line are configured for local overreach distance protection \cite{tamronglak_analysis_1994, elizondo_methodology_2003, blackburn_protective_2007}.
The PUTT relaying scheme is illustrated in Fig.~\ref{fig:putt-oneline}.
In order to trip, the relays must receive a remote enable signal from the other end of the line, where the remote signal is based on the output of an underreaching distance relay.
As is the case for DCUB, there are two unidirectional communication channels that broadcast a continuous blocking tone.
This scheme can experience hidden failures when: either or both of the underreaching units continuously report a fault; or either or both of the receivers continuously report an unblocking signal.

\noindent \textit{Permissive Overreaching Transfer Trip (POTT) Relays}
\vspace{0.1in}
\indent

\begin{figure}[!htbp]
\centering
\includegraphics[scale=0.225]{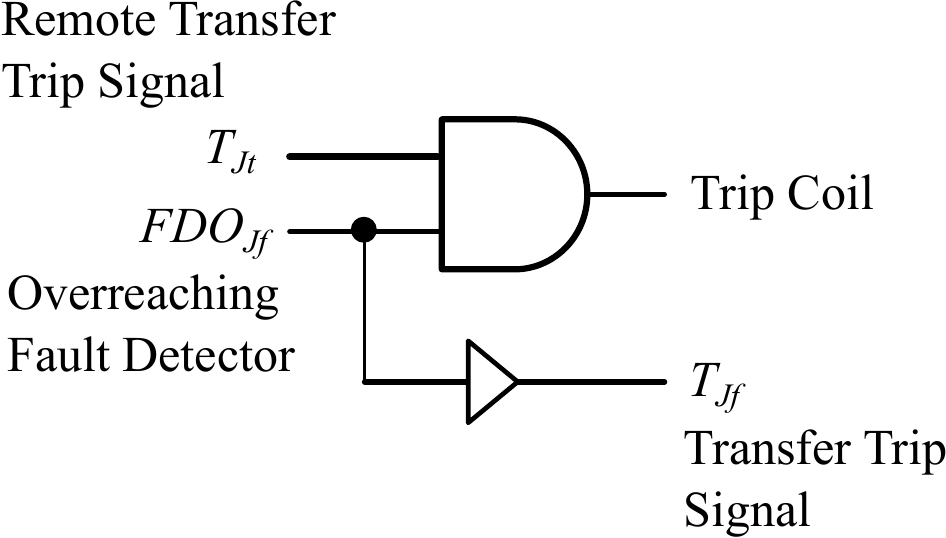}
\caption{Schematic for one relay in a POTT scheme.}
\label{fig:pott-logic}
\end{figure}

In a POTT relaying scheme, the protective relays on either end of the line are again configured for local overreach distance protection \cite{tamronglak_analysis_1994, elizondo_methodology_2003, blackburn_protective_2007}.
In this case, however, there is not a separate underreaching remote relay, so the overreaching distance element on the far end of the line is used to produce the unblocking signal, as illustrated in Fig.~\ref{fig:dcub-oneline}. A schematic for one such relay in the oneline diagram from Fig.~\ref{fig:dcub-oneline} is illustrated in Fig.~\ref{fig:pott-logic}.
This scheme can experience hidden failures for similar reasons as the PUTT relaying scheme:
either \emph{but not both} of the underreaching unit continuously report a fault; or either or both of the receivers continuously report an unblocking signal.

\vspace{0.1in}
\noindent \textit{Zone~1 / Zone~2 / Zone~3 Distance Relays}
\vspace{0.1in}
\indent

\begin{figure*}[!htbp]
\centering
\includegraphics[scale=0.5,angle=0]{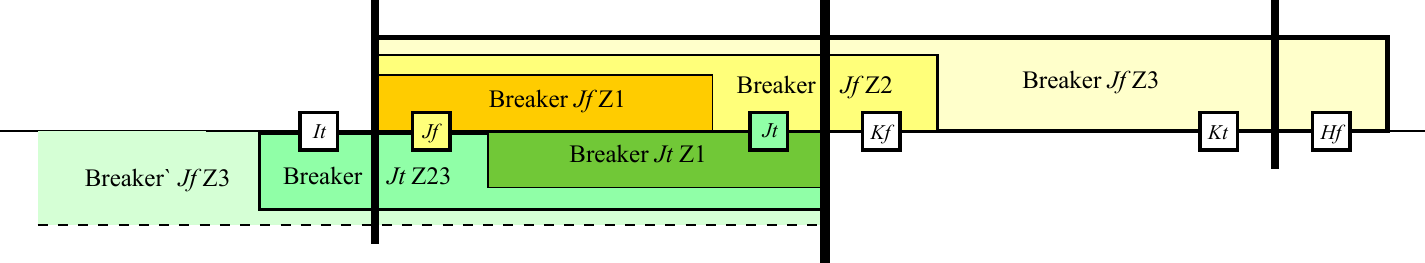}
\caption{Oneline diagram for two relays in a Zone~1 / Zone~2 / Zone~3 protection scheme.}
\label{fig:z123-oneline}
\end{figure*}

For the case of lines Zone~1 / Zone~2 / Zone~3 distance protection, illustrated in Fig.~\ref{fig:z123-oneline}, the Zone~2 and Zone~3 relays can experience hidden failures should the delay timers (T2 and T3 in the figure) fail in the ``closed'' position \cite{tamronglak_analysis_1994, elizondo_methodology_2003}.
In the event of a Zone~2 or Zone~3 fault on an adjacent line, the relay will operate instantaneously for a fault resulting in the line being incorrectly removed from service.

\vspace{0.1in}
\noindent \textit{Phase Comparison Blocking Relays}
\vspace{0.1in}
\indent

Phase comparison blocking relays are an approximation of differential relays whereby relays compare a local and remote signal to determine if the phase difference between the two signals is less than a threshold \cite{tamronglak_analysis_1994, elizondo_methodology_2003}.
Typically, the signal measured is the negative sequence current flowing into a transmission line, and the transmitted signal is its sign.

\begin{figure}[!htbp]
\centering
\includegraphics[scale=0.6]{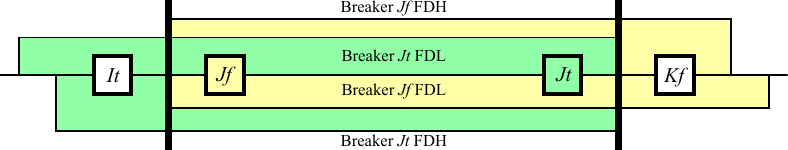}
\caption{Oneline diagram for a phase comparison scheme.}
\label{fig:phase-comparison-oneline}
\end{figure}

This relaying typically uses a pair of fault detectors: a ``high'' fault detector to enable local tripping and a ``low'' fault detector to enable the transmission. The ``high'' fault detector must be less sensitive than the ``low'' fault detector to avoid tripping for out of zone faults, as illustrated in  Fig.~\ref{fig:phase-comparison-oneline}.
The system is designed so that it will operate as an overcurrent relay if the phase comparison signal is not received, potentially resulting in out-of-zone trips. This can occur when: the communications link has failed; either or both ``high'' fault detectors continuously report a fault; or either or both ``low'' fault detectors fail to detect a fault. It is important to note that if the ``high'' fault detectors fail concurrently with either channel, an immediate trip will result.

\vspace{0.1in}
\noindent \textit{Directional Overcurrent Ground Relays}
\vspace{0.1in}
\indent

Directional overcurrent ground relays determine direction based on the difference in angle between phase currents and a reference quantity, referred to as the polarizing source. Typically this reference quantity is the zero-sequence current flowing through a transformer grounding point, though sometimes the zero-sequence voltage at a transformer grounding point is used \cite{blackburn_protective_2007}.
These relays can trip for a backward (out-of-zone) fault if the relay loses directionality derived from the polarizing source \cite{tamronglak_analysis_1994}.

\vspace{0.1in}
\noindent \textit{Differential Relays}
\vspace{0.1in}
\indent

Differential relays monitor the currents flowing into each terminal of the protected element and trip if the sum of these currents differs significantly from zero. Restraint functionality prevents tripping resulting from slight mismatches in current transformer parameters if the protected element is operating at high power or from inrush current \cite{blackburn_protective_2007}. This relaying is typically used for protecting transformers, buses, and generators.
A traditional electromechanical relay using restraint coils can experience a hidden failure if the restraint coils are shorted, resulting in a misoperation during high load, transformer energization, or external faults \cite{tamronglak_analysis_1994, elizondo_methodology_2003}.

\vspace{0.1in}
\noindent \textit{Breaker Failure Relays}
\vspace{0.1in}
\indent

Circuit breakers typically have output contacts to indicate their state to the protection system. Should these contacts be shorted or a short occur in the measurement lines, the protective relaying will believe that the breaker failed to operate, potentially resulting in an erroneous transfer trip occurring \cite{tamronglak_analysis_1994,elizondo_methodology_2003}.

\vspace{0.1in}
\noindent \textit{Impact on Digital Relays}
\vspace{0.1in}
\indent

For modern integrated digital relays, the likelihood of some failure modes is reduced.
In the case of DCB, it is unlikely that the reverse reach fault detectors will fail to detect a fault while the forward fault detection works correctly.
For Zone~2 and Zone~3 protection, there are no mechanical timers to fail, while for differential protection, there is no separate restraint winding to short.

Additionally, for digital relays with monitoring capability many hidden failures can be detected or mitigated. For example, a supervisory system can monitor for: continuously received unblocking signal, the lack of signal from a current transformer or voltage transformer (caused by a short or open circuit), and reversed wiring for a current transformer or voltage transformer that is indicated by high negative-sequence quantities.
For the case of distance protection, integrated digital relays makes it more straightforward to include an overcurrent detector that must trip for distance protection to activate, preventing a breaker from operating due to a lost signal from a voltage transformer.

However, digital relays with supervisory systems still will not detect some hardware failures -- e.g., shorted circuit breaker output contacts or a shorted current transformer on a transformer neutral when the system is close to balanced -- and will not be able to tell if an improper protection scheme is applied or protection settings are incorrect -- e.g. the operating area for distance protection is too large or restraint slope for differential protection is too low -- for a given scheme.

\section{Identifying Critical Relays} \label{sec:identifying}

\vspace{0.05in}
\noindent \textit{Methods for Identifying Critical Relays}
\vspace{0.1in}
\indent

A number of methods have been developed to identify critical relays in an electrical grid in terms of their role in causing or propagating cascading failures.
Software tools for identifying critical locations for cascading failures fall into three main categories: linearized methods, static simulation-based methods, and dynamic simulation-based methods \cite{baldick_initial_2008, papic_survey_2011}. A fourth category would be graph-based or topological methods, which are based on models of the spread of disease epidemics, however, research has shown that results from such methods do not correlate well with those based on the physics of electric power flow \cite{hines_topological_2010}.
Criteria for predictive tools for cascading failures include: accuracy of reproduction of the physical system, speed of execution, requirement for large amounts of input data, ability to review and explain results, limits on the size of the modeled network, and method for solving the network equations (e.g., ac analysis, dc approximation, relaxation) \cite{vaiman_2011}.

Linearized methods are also referred to as small signal assessment (SSA) tools \cite{baldick_initial_2008}. These are simple to implement, computationally efficient, and are included in commercial tools such as PowerWorld\textsuperscript{\textregistered} and GE's Positive Sequence Load Flow (PSLF) software. However, the linear approximations used can introduce errors when identifying contingency sets that produce cascading outages \cite{yang_comprehensive_2007, eppstein_random_2012}.

Static simulation methods rely on assessing a potential contingency -- selected from a range of possible contingency scenarios using power flow methods -- with a steady-state power flow solution.
The potential for a cascading event can be determined based on the output of the power flow \cite{phadke_expose_1996} using criteria such as: 1) the normalized change in the determinant of the Jacobian matrix from the base case; 2) the line loading; 3) the largest deviation of the phase angle in the first iteration of a Fast Decoupled Load Flow following the contingency, and using the base case solution as an initial point; 4) the Gaussian norm of the deviation of all bus voltage angles after the first iteration. As some of these criteria rely on a single iteration, those could be considered linear methods.
Alternatively, a minimum load shed (MLS) formulation \cite{yang_effects_2006} can be used: an MLS approach is guaranteed to provide a solution, but it will not capture the effects of cascading failures resulting from hidden failures.

Transient simulation methods rely on a time-domain simulation, which can vary from a series of static power flows to a true dynamic simulation (i.e., modeling the generator rotors and prime movers, which allows the failure mechanism of generator pole slipping to be modeled).
A handful of software tools are available:
DCSIMSEP is based on a series of dc power flows \cite{eppstein_random_2012}.
TRELSS is based on a series of steady-state ac power flows \cite{bialek_2016}.
COSMIC models the mechanical and controller dynamic behavior of generators \cite{song_2016}.
PowerWorld\textsuperscript{\textregistered} includes a transient dynamics module that can model the dynamic behavior of a power system, including relay behavior.

\vspace{0.1in}
\noindent \textit{Challenges of Identifying Critical Relays}
\vspace{0.1in}
\indent

Even though software tools for identifying critical locations for cascading failures based on linear approximations are computationally efficient and are able to handle combinations of asset failures for networks of practical size, such tools are limited in accuracy by the linearized approximations of a highly nonlinear system, which may lead to mischaracterization in terms of critical lines and substations \cite{yang_effects_2006}.

Despite the computational and data challenges described, significant progress has been made in terms of applying simulation-based tools to identify contingency sets (either branches or buses).
Hines~et~al.~\cite{hines_dual_2013} describes how the random chemistry method may be applied to find collections of contingencies that lead to cascading failures.
It is important to distinguish cascading contingency analysis from the $n-k$ interdiction problem \cite{sundar_probabilistic_2018}, where the objective is to find the worst set $k$ assets in terms of maximizing load loss, where $k$ is a small number.

Many types of interactions between power system assets can propagate a cascading failure: transmission line tripping causing a frequency transient or overloading of other lines, operation or misoperation of relays, reactive power problems, system instabilities, or operator stress \cite{nedic_2006}.
Fitzmaurice~et~al.~\cite{fitzmaurice_2012} claim that there is a high correlation between the sequences of branch failures in cascading events between dynamic and quasi-steady-state models, while topological models show low correlation compared to the other two.

Simulating cascading failures is challenging due to a number of reasons:
It involves the simulation of a problem with processes that occur on radically different time scales, namely the response time of protective relaying and the mechanical time constants of generation.
Many different types of interactions between assets that propagate cascading events need to be modeled.
The operation of protective relaying introduces discontinuities in the solutions of the set of differential equations governing the system and it may not be possible for the solver to find a solution after a relay operation; for this reason, commercial tools for transient simulation often neglect the behavior of protection.
Lastly, the data requirements for cascading failure modeling can be quite high, particularly if dynamic modeling is performed; potentially required data includes generator cost or dispatch constraints, power flow limits, protection system and relay data, branch outage probabilities, breaker failure probabilities, bus fault probabilities, node-breaker topology, loads, hazards, etc. \cite{bialek_2016}.

Of particular importance is the node-breaker topology information, as bus-branch representations used for power flow studies do not capture substation breaker configurations.
One factor that does help with the use of cascading failure analysis in the context of selecting sets of critical assets for hidden failures is that hidden failures will only cause misoperations within a ``region of vulnerability'' that contains the external fault, where the region is typically restricted to adjacent lines and substations \cite{qiu_risk_2003, tamronglak_analysis_1994}.

\section{Mitigation of Hidden Failures} \label{sec:mitigation}
\indent

Digital relays offer some protection against hidden failures through their ability to perform self-tests. That capability, however, is limited in that it cannot detect errors on both the input (i.e., analog signal conditioning section) and the output contactor sections, nor can it detect the validity of the settings.
Maintenance is necessary to detect errors in these cases, however, there is evidence that maintenance becomes the main source of hidden failures as a relay technician may fail to restore the original settings \cite{gao_case_2013}.

In case a substation is identified as critical, several mitigation methods are available to reduce or eliminate the likelihood of a hidden failure occurring.
Such methods include using: relays for local backup or redundant protection systems, supervisory systems, voting schemes, adaptive protection, and centralized protection. These are discussed in detail below.

\vspace{0.1in}
\noindent \textit{Relays for Backup}
\vspace{0.1in}
\indent

Using identical relay models or relays made by the same manufacturer is recommended and an ongoing trend in utility practice \cite{sandoval_using_2010}.
In the past, many utilities installed relays from different manufacturers as local backup or for redundant systems with the theory being that failures between relays of different manufacturers are more likely to be uncorrelated; some utilities continue to do so in the present day. However, the aviation industry has demonstrated the ability to build high-reliability systems with \emph{identical} redundant components.
In addition, internal components from relays tend to use the same vendors so the benefit of failure independence is reduced. Most important, however, is the fact that using different manufacturers increases the labor involved in configuring settings and thereby increases the likelihood of improper settings and a misoperation in a redundant configuration.

\vspace{0.1in}
\noindent \textit{Supervisory Systems}
\vspace{0.1in}
\indent

Adding a supervisory system to detect whether a relay has a hidden failure is discussed by Gao~et~al.~\cite{gao_case_2013}.
Their system relies on the relay having IEEE~61850 or similar communications capabilities and the presence of additional measurement devices (e.g., metering).
If measurement devices share the same voltage or current transducers, the supervisory system may not be able to detect failures in the transducers, but will be able to detect failures in the analog front ends of either the metering or protection devices. The system may also be able to detect misconfigured settings by recording internal states of the protective relaying and measurements from other sources when a fault or other event occurs on the system.

Alternatively, a different supervisory system is proposed Phadke~et~al.~\cite{phadke_expose_1996} to prevent relays from operating when a hidden failure occurs. Their system monitors the relay input signals and provides output contacts that are connected in series with the relay output contacts. Therefore the supervisory system behaves as a similar relay connected in a logical ``and'' with the existing relay.

\vspace{0.1in}
\noindent \textit{Voting Schemes}
\vspace{0.1in}
\indent

Using voting schemes to reduce hidden failures by increasing the security of relays is discussed in \cite{sandoval_using_2010} and \cite{nye_liu_hofmann_2005}. 
Sandoval~et~al.~\cite{sandoval_using_2010} demonstrated, by relying on fault tree analysis, that a two-out-of-three voting scheme reduces the probability of a misoperation for a transformer protection scheme by a factor of three.
However, voting schemes do not eliminate all misoperations in the case of hidden failures, as discussed by Albinali~et~al.~\cite{albinali_2016}.

\newpage
\noindent \textit{Adaptive Protection}
\vspace{0.1in}
\indent

Adaptive protection is a concept by which digital relays change their settings over time in response to local or remote measurements on an electrical grid \cite{horowitz_adaptive_1988}, but still operate based only on local measurements.
The motivation behind adaptive protection is to allow relays to change their settings so they maintain a reasonable trade-off between security and dependability under varying conditions. These conditions include: voltage magnitude or frequency variations, changes in switching states (e.g., connecting or disconnecting line tap, changing transformer tap ratio), nonlinear behaviors (e.g., transformer saturation during inrush), and varying load flows on parallel transmission lines.

Using adaptive protection to reduce the likelihood of hidden failures by adjusting relay settings to prevent their operation during transient events (e.g., cold-load pickup) is discussed by Horowitz~et~al.~\cite{horowitz_boosting_2003}.
Even though adaptive protection methods are able to reduce the potential for misoperation with incorrect settings or contingency conditions (e.g., voltage instability), as discussed by Jonsson~et~al.~\cite{jonsson_adaptive_2003}, such methods will not help to protect against hidden failures in measurement components.
Failures in measurement components can occur in the transducer itself (voltage or current transformer), measurement line, or the analog front end of a relay. These errors cannot be detected via digital relay self-tests, but it is possible to detect and correct for them with centralized protection.

\vspace{0.1in}
\noindent \textit{Centralized Protection}
\vspace{0.1in}
\indent

Centralized protection extends the concept of adaptive protection via replacing individual relays with centralized control in which breakers operate based on multiple (both local and remote) measurements.
A centralized protection scheme, which uses dynamic state estimation to both detect hidden failures and allow protection to operate without performance degradation after such a failure has been detected, is proposed by Albinali~et~al.~\cite{albinali_2016}, based on the method developed by Meliopoulos~et~al.~\cite{meliopoulos_settingless_2013}. It makes use of local measurements at a substation except for transmission line protection, in which case measurements must be received from the other end of the protected line. Furthermore, it makes use of intelligent electronic devices (IEDs) located near current transformers and voltage transformers (outside of the control house) so measurements are sent via IEEE~61850 over a fiber-optic network.
The measurements for an entire substation are collected by a dynamic state estimator, which creates a dynamic model of the substation. For each measurement, a chi-squared goodness-of-fit test is applied to determine whether there is a measurement hidden failure; if a measurement fails the test, it is removed and the state estimation is repeated.
The method of Meliopoulos~et~al.~\cite{meliopoulos_settingless_2013} has been developed to support the protection of substation transformers \cite{fan_dynamic_2015, meliopoulos_dynamic_2017}, transmission lines \cite{meliopoulos_settingless_2013}, and shunt capacitor banks \cite{meliopoulos_settingless_2013}.

Protection of transmission lines presents a challenge as the method of Meliopoulos~et~al.~\cite{meliopoulos_settingless_2013} needs high-speed communication between substations or switching stations at both ends of the line; it requires real-time measurements of the voltage and current phasors at each phase on each end of a transmission line.
Additionally, Meliopoulos~et~al. propose to extend their method to detect misoperations from relay hidden failures by representing the substation in terms of a hierarchical organization and detect relay misoperation in case the region where a relay operates corresponds with a detected abnormality by the centralized protection system.

Other centralized protection methods have been proposed that make use of measurements throughout a transmission network.
Neyestanaki~et~al.~\cite{neyestanaki_adaptive_2015} propose a method based on distributed phasor measurements to detect faults and the locations of their positions on transmission lines; the number of phasor measurement unit (PMU) locations required for the method to operate, where a PMU is not required at every bus, is discussed.

\section{Conclusions} \label{sec:conclusion}
\indent

Outages caused by hidden failures are rare compared to those caused by natural disasters (e.g., extreme weather, forest fires, earthquakes). However, research into cascading failures in electrical grids has demonstrated that there are relatively small collections of critical assets that can cause large outages via cascading failures or load shedding from remedial action schemes. The set of assets disabled by a hidden failure can contain one of these collections of critical assets resulting in a large outage as has occurred historically.
Given that the number of critical assets in an electrical grid is relatively low, investment in protection system upgrades to reduce the likelihood of hidden failures may likely be cost-effective compared with investments in storm-hardening transmission network infrastructure.

The expense in reducing the likelihood of hidden failures may not be high, as several ``hidden'' failure modes can in fact be detected \emph{directly} with monitoring, especially those associated with pilot protection.
Other failure modes (e.g., shorted timers) are unlikely to occur for integrated digital relays, so the practice of electromechanical relays for backup protection may be inadvisable for critical locations where security is important.
For some hidden failures that cannot be detected directly, state-based centralized protection offers a potential solution, especially for applications with nonlinear elements or highly varying operating conditions (e.g., transformer protection or tapped line protection).
Other design philosophies can help mitigate the likelihood of misoperations: this includes the use of identical systems for redundancy and the preference for local over remote backup protection.


\bibliographystyle{unsrt}
\bibliography{references}

\end{document}